\documentclass[12pt,a4paper]{article}
\usepackage{amsfonts, amsmath, amssymb}
\usepackage{graphicx,xspace,psfrag,color}  
\usepackage{mathrsfs}
\usepackage{geometry,hyperref,latexsym,pifont,textcomp,setspace}
\usepackage{bm}
\newcommand{\scri}{\mathscr{I}}

\psfrag{t}[][]{\Large$t$}
\psfrag{r}[][]{\Large$r$}
\psfrag{ip}[][]{$i^+$}
\psfrag{im}[][]{$i^-$}
\psfrag{i0}[][]{$i^0$}
\psfrag{scrip}[][]{$\scri^+$}
\psfrag{scrim}[][]{$\scri^-$}
\title{\Huge\bf Uncovering the effective spacetime$^*$ \linebreak\linebreak\normalsize- Lessons from the effective field theory rationale -}

\author{
Carlos Barcel\'{o}$^1$,
Ra\'ul Carballo-Rubio$^{1}$ and
Luis J. Garay$^{2,3}$\\
\small\it $^1$ Instituto de Astrof\'{i}sica de Andaluc\'{i}a, CSIC,\\
\small\it Camino Bajo de Hu\'{e}tor 50, 18008 Granada, Spain\\
\small\it $^2$  Departamento de F\'{i}sica Te\'{o}rica II,
Universidad Complutense de Madrid,\\
\small\it 28040 Madrid, Spain\\
\small\it $^3$  Instituto de Estructura de la Materia, CSIC,\\
\small\it Serrano 121, 28006 Madrid, Spain
}

\def\fecha{2015/03/31}

\begin{document}
\maketitle
\thispagestyle{empty}


\bigskip
\hrule
\begin{abstract}
\noindent
The cosmological constant problem can be understood as the failure of the decoupling principle behind effective field theory, so that some quantities in the low-energy theory are extremely sensitive to the high-energy properties. While this reflects the genuine character of the cosmological constant, finding an adequate effective field theory framework which avoids this naturalness problem may represent a step forward to understand nature. Following this intuition, we consider a minimal modification of the structure of general relativity which as an effective theory permits to work consistently at low energies, i.e., below the quantum gravity scale. This effective description preserves the classical phenomenology of general relativity and the particle spectrum of the standard model, at the price of changing our conceptual and mathematical picture of spacetime.

\end{abstract}
\hrule
\bigskip

\maketitle
 \begin{center}
$^*$ This essay received an Honorable Mention in the 2015 Essay Competition of the Gravity Research
Foundation.
 \end{center}
\vfill

\bigskip
{\underline{E-mail}: carlos@iaa.es, raulc@iaa.es, luisj.garay@ucm.es}

\clearpage
\markright{Uncovering ... \fecha \hfil }
\pagestyle{myheadings}

\def\HRULE{{\bigskip\hrule\bigskip}}




Oh no, yet another discussion about the cosmological constant problem! Yeap, that it is. When one listens so many times that this is one of the most severe and fundamental problems faced by modern theoretical physics, it is difficult to resist the temptation of having a look by yourself to see what is going on ... and many of us do! One is eager to have his own reading of at least the main arguments involved.

Within the framework of effective field theory the cosmological constant term corresponds to a relevant operator that is, however, not natural \cite{Polchinsk1992,Burgess2013}. Therefore the value of the cosmological constant is highly sensitive to the ultraviolet details beyond the effective theory, and it has to be fine-tuned in order to match the experiments. On the one hand, this observation is compelling: the value of the cosmological constant is an issue to be treated in a theory which consistently unifies the ultraviolet and infrared details of our universe, that is, a theory of quantum gravity. On the other hand, it challenges the basic working principle according to which the behavior of physics at a given distance scale is insensitive to the fine details of the dynamics at much shorter distances. In our opinion, it would be desirable that an effective theory rationale exists that permits to work consistently at low energies, leaving the question about the value of the cosmological constant unanswered until we are able to construct a more complete theory. From the perspective of  low-energy physics, in such a framework the cosmological constant will be as mysterious as (but not more than) any other parameter in physics such as the gravitational constant or the electron charge. 

The naturalness problem we have delineated is of ultraviolet nature. Indeed, it is enough to consider an effective description of the physics of the solar system to demonstrate its existence \cite{Martin2012,Kagramanova2006}: solar system experiments constrain the possible cosmological constant as $|\Lambda|\leq 10^{-32}\mbox{ GeV}^4$. This value is far smaller than any result coming from straightforward theoretical calculations, which is roughly $|\Lambda_{\mbox{\tiny vac}}|=10^{8}\mbox{ GeV}^4$. This should not be mixed, as it happens sometimes in the literature, with the problem of explaining the observed value of the cosmological constant (and related issues as, e.g., the cosmic coincidence problem). To explain this value one would have to consider a much wider range of scales, from very short lengths outside the domain of applicability of the effective field theory up to the Hubble length. Let us stress that the need of taking into account so large length scales does \emph{not} make the problem bona fide infrared. 

Our guiding principle is to consider the minimal deviations from known physics which permit to save the effective field theory rationale, so we require the following tight constraints. First, given the quantum nature of the problem, modifications of the classical physics lack of any compelling motivation and the content of classical general relativity will be largely respected. Second, the problem has its roots in the coupling to the gravitational field, so that the particle spectrum of the standard model will be preserved. In our framework there are no extra degrees or freedom or dimensions, but rather a different view of spacetime. In this sense it can be understood as a minimal approach to the problem, making it genuinely different from other approaches.

The problematic cosmological constant term in general relativity is of the form $\sqrt{|g|}\,\Lambda$. Naively one can envision that, if only there existed a background volume form $\bm{\omega}$ (with corresponding density $|\omega|$), one could 
replace the cosmological constant term by $\sqrt{|\omega|}\,\Lambda$. This term, corresponding to a constant shift in the Lagrangian density, would then be irrelevant to the classical dynamics. However, following the effective field theory logic, this condition alone would not be enough: a symmetry that forbids the occurrence of $\sqrt{|g|}\,\Lambda$ in the Lagrangian density must be present in order to guarantee that this term is not generated by radiative corrections. A natural candidate for the job is the invariance under scale transformations of the gravitational field,
\begin{equation}
g_{ab}\rightarrow \zeta^2g_{ab},\qquad\qquad \zeta\in\mathbb{R}.\label{eq:scale}
\end{equation}
A consistent and well-motivated gravitational theory incorporating this symmetry requirement is known as Weyl transverse gravity \cite{Izawa1995,Alvarez2006}. Notice that the invariance under longitudinal diffeomorphisms of general relativity would be broken to guarantee the invariance under \eqref{eq:scale} while keeping us in a second-order field theory. In order to maintain the number of degrees of freedom in the gravitational sector, the symmetry \eqref{eq:scale} is then extended to a gauge symmetry. That is, in Weyl transverse gravity the gravitational symmetries are given by transverse diffeomorphisms and Weyl transformations. This theory indeed corresponds to the only alternative way of defining a nonlinear theory of the two local degrees of freedom associated with gravitational waves \cite{Barcelo2014,BCRG2014}.

The nature of the spacetime associated with these symmetries can be condensed in the notion of a differentiable manifold equipped with a conformal structure (see, e.g., \cite{Ehlers2012}) and a non-dynamical volume form. From a more physical perspective, we could say that in Weyl transverse gravity spacetime is an incompressible continuum. The inclusion of matter in the picture has to preserve the gravitational symmetries by, e.g., ensuring that matter is inert under Weyl transformations. This permits to accommodate a general matter content and, in particular, the standard model of particle physics, thus satisfying one of our requirements.

Despite this departure from the principles behind general relativity, this gravitational theory turns out to be equivalent to the former at the classical level, in the following sense: there exists a common gauge choice in which the field equations in the presence of a general classical matter take the same form \cite{Alvarez2006,Ellis2011,Ellis2014}. The cosmological constant appears as an integration constant, unrelated to any coupling in the gravitational action. This equivalence fulfills the condition of preservation of the phenomenology of general relativity, by retaining the space of solutions of the classical field equations. The importance of this apparently simple observation should not be underestimated: all in all, the classic tests of general relativity \emph{cannot be used} to argue that spacetime possesses a metric volume element $\text{d}^4x\sqrt{|g|}$.

Classical equivalence in this sense does not imply the equivalence of the corresponding quantum theories, as it is well known. Quantization is a subtle procedure; even for systems with a finite number of degrees of freedom one easily finds unexpected results \cite{Groenewold1946}. Even more surprising, quantization and gauge fixing do not generally commute \cite{Loll1990}. These results show how counterintuitive the procedure of quantization may result.

The analysis of the semiclassical model in which matter fields are quantized but the gravitational field is treated classically turns out to be enough to display explicitly the differences. By construction, in Weyl transverse gravity the cosmological constant problem is avoided: the symmetry \eqref{eq:scale} protects the cosmological constant sector against radiative corrections, provided that it can be implemented in the presence of quantum effects. This symmetry is known to be anomalous when general conditions are met, and one of these general conditions is invariance under the \emph{full} group of diffeomorphisms. An anomaly is an inconsistency between different symmetries; in the case at hand, it is a consequence of the interplay between longitudinal diffeomorphisms and Weyl transformations. When the longitudinal diffeomorphisms are left out of the game, as it is precisely the case in Weyl transverse gravity, classic results on scale-invariance anomalies are circumvented, guaranteeing that gravitational scale invariance survives quantum corrections \cite{Carballo-Rubio2015}.

Our discussion motivates the question about the reality of gauge symmetries. Gauge symmetries are not directly measurable, as only gauge-invariant quantities are. Thus we can only indirectly infer their existence by analyzing the corresponding implications. While Weyl transverse gravity is equivalent, in a sort of degeneracy, to general relativity at the classical level, quantum effects do discriminate between these theories. In particular, in Weyl transverse gravity the low-energy effective theory is self-consistent, with no low-energy parameters requiring fine-tuning. Any value of the cosmological constant inserted in the classical field equations will be automatically stable under radiative corrections.

At the end of the day, we see that it is possible to maintain the effective field theory rationale. That a different but simple vision of spacetime is enough to guarantee this is conceptually compelling. We can say that, from an honest low-energy perspective and combining known theories and experiments, there is a strong case in favor of the alternative gauge structure of the gravitational interaction against diffeomorphism invariance; that is, in favor of a picture of spacetime as a continuum with zero compressibility, or equipped with a conformal structure and a non-dynamical volume form in mathematical terms. Regarding the cosmological constant, only a more fundamental theory or principle could unveil its nature and set its actual value, which by matching should be the one used at low energies (in particular, this principle may be related to the \emph{true} vacuum of the high-energy theory; see, e.g., Volovik's proposal~\cite{Volovik2009}). It is the task of future research to examine the potential role of these low-energy properties in the way towards a more fundamental description of the gravitational interaction.

\vspace{0.5cm}

\textbf{Acknowledgments:}
Financial support was provided by the Spanish MICINN through Projects No. FIS2011-30145-C03-01 and No. FIS2011-30145-C03-02 (with FEDER contribution), and by the Junta de Andaluc\'ia through Project No. FQM219. R. C-R. acknowledges support from CSIC through the JAE-predoc program, cofunded by FSE.

\bibliography{grf2015}	

\begin{thebibliography}{10}

\bibitem{Polchinsk1992}
J.~{Polchinski}.
\newblock {Effective Field Theory and the Fermi Surface}.
\newblock {\em ArXiv e-prints}, October 1992.

\bibitem{Burgess2013}
C.~P. {Burgess}.
\newblock {The Cosmological Constant Problem: Why it's hard to get Dark Energy
  from Micro-physics}.
\newblock {\em ArXiv e-prints}, September 2013.

\bibitem{Martin2012}
J.~{Martin}.
\newblock {Everything you always wanted to know about the cosmological constant
  problem (but were afraid to ask)}.
\newblock {\em Comptes Rendus Physique}, 13:566--665, July 2012.

\bibitem{Kagramanova2006}
V.~{Kagramanova}, J.~{Kunz}, and C.~{L{\"a}mmerzahl}.
\newblock {Solar system effects in Schwarzschild de Sitter space time}.
\newblock {\em Physics Letters B}, 634:465--470, March 2006.

\bibitem{Izawa1995}
K.~{Izawa}.
\newblock {Derivative Expansion in Quantum Theory of Gravitation}.
\newblock {\em Progress of Theoretical Physics}, 93:615--619, March 1995.

\bibitem{Alvarez2006}
E.~{{\'A}lvarez}, D.~{Blas}, J.~{Garriga}, and E.~{Verdaguer}.
\newblock {Transverse Fierz Pauli symmetry}.
\newblock {\em Nuclear Physics B}, 756:148--170, November 2006.

\bibitem{Barcelo2014}
C.~Barcel{\'o}, R.~Carballo-Rubio, and L.~J. Garay.
\newblock {Unimodular gravity and general relativity from graviton
  self-interactions}.
\newblock {\em Phys. Rev.}, D89:124019, June 2014.

\bibitem{BCRG2014}
C.~{Barcel{\'o}}, R.~{Carballo-Rubio}, and L.~J. {Garay}.
\newblock {Absence of cosmological constant problem in special relativistic
  field theory of gravity}.
\newblock {\em ArXiv e-prints}, June 2014.

\bibitem{Ehlers2012}
J.~Ehlers, F.~A.~E. Pirani, and A.~Schild.
\newblock Republication of: The geometry of free fall and light propagation.
\newblock {\em General Relativity and Gravitation}, 44(6):1587--1609, June
  2012.

\bibitem{Ellis2011}
G.~F.~R. {Ellis}, H.~{van Elst}, J.~{Murugan}, and J.-P. {Uzan}.
\newblock {On the trace-free Einstein equations as a viable alternative to
  general relativity}.
\newblock {\em Classical and Quantum Gravity}, 28(22):225007, November 2011.

\bibitem{Ellis2014}
G.~F.~R. {Ellis}.
\newblock {The trace-free Einstein equations and inflation}.
\newblock {\em General Relativity and Gravitation}, 46:1619, January 2014.

\bibitem{Groenewold1946}
H.~J. {Groenewold}.
\newblock {On the principles of elementary quantum mechanics}.
\newblock {\em Physica}, 12:405--460, October 1946.

\bibitem{Loll1990}
R.~{Loll}.
\newblock {Noncommutativity of constraining and quantizing: A U(1)-gauge
  model}.
\newblock {\em Phys. Rev. D}, 41:3785--3791, June 1990.

\bibitem{Carballo-Rubio2015}
R.~{Carballo-Rubio}.
\newblock {Longitudinal diffeomorphisms obstruct the protection of vacuum
  energy}.
\newblock {\em ArXiv e-prints}, February 2015.

\bibitem{Volovik2009}
G.~E. Volovik.
\newblock {\em The Universe in a Helium Droplet}.
\newblock International Series of Monographs on Physics. OUP Oxford, May 2009.

\end{thebibliography}

\end{document}